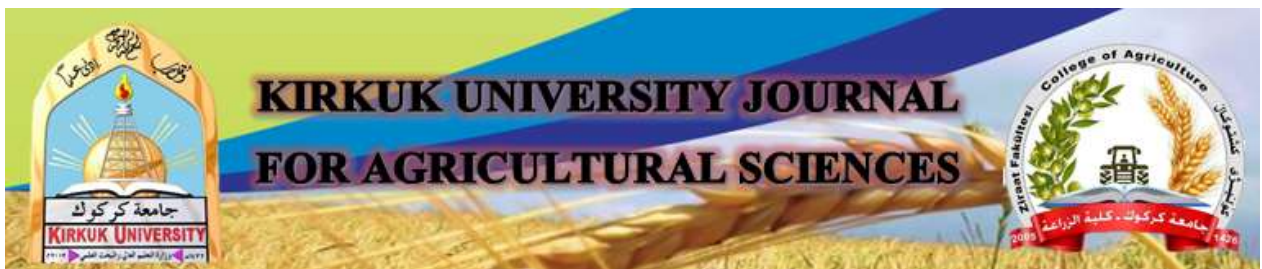

# Rooting of thornless blackberry cuttings as induced by the extract of white willow (*Salix alba* L.) shoots collected in different times

Kocher Omer Salih[1]  
kocher.salih@univsul.edu.iq

Aram Akram Mohammed[2]  
aram.hamarashed@univsul.edu.iq

Ibrahim Maaroof Noori[3]  
ibrahim.nwri@univsul.edu.iq

[1, 2, 3] Horticulture Department, College of Agricultural Engineering Sciences, University of Sulaimani, Sulaymaniyah, Kurdistan Region, Iraq.

## Abstract

The aqueous extract of *Salix spp* contains many compounds which may act as root-promoting agents in cuttings. *S. alba* is a deciduous tree containing variable phytochemicals which are variable throughout the year. So, in this study, one- and two-year-old shoots of *S. alba* were collected on the 15$^{th}$ of each month in the year 2022, extracted in 2% ethanol at 9 g.L$^{-1}$, and placed in a water bath at 35 °C, then they applied to thornless blackberry cuttings for 1.5 hr. The results explained that the highest rooting percentage (66.67%) was obtained in the cuttings soaked in the extract of willow shoots collected on 15$^{th}$ of January. They were not significantly different from control cuttings, but they were different from the cuttings soaked in the extract of willow shoots collected on 15$^{th}$ of August and October (33.33%). The majority of other shoot and root traits were high in the cuttings soaked in the extract of willow shoots collected on 15$^{th}$ of December. The willow shoots collected on 15$^{th}$ of January contained the lowest total phenols (51.4 µg.mL$^{-1}$) and total flavonoids (29.07 µg.mL$^{-1}$). Moreover, the highest total phenols (57 µg.mL$^{-1}$) and IAA (365.17 µg.mL$^{-1}$) were recorded in the willow shoots collected on 15$^{th}$ of March, however each total flavonoids (44.96 µg.mL$^{-1}$) and salicylic acid (492.61 µg.mL$^{-1}$) were the highest in the willow shoots collected on 15$^{th}$ of April. Generally, based on rooting percentage, it is advisable to collect willow shoots on 15$^{th}$ of January and February for extraction and application to the thornless blackberry cuttings.

**Introduction**

Since the findings obtained by [1], it has been approved that auxin; indole acetic acid (IAA) has a crucial role in adventitious root formation in cuttings. They also found that exogenous synthetic IAA had the same effect when applied to the cuttings. However, auxin has not been admitted as a single agent which is responsible for rooting in the cuttings, because many other substances may interact with auxins to induce root formation, such as sugars, phenolic compounds, and oxidase enzymes [2]. Therefore, many attempts have been exerted to find natural sources of auxins and other substances that promote root formation in cuttings with lower costs and lesser harms to humans and the ecosystem [3]. Among the most popular natural extracts that are frequently used for cuttings to improve rooting are plant extracts, particularly white willow (*Salix alba* L.) extract.

White willow is a deciduous tree from the Salicaceae family growing up to 30 m. It is endemic to Europe, Asia, and North Africa [4, 5]. The aqueous extract of *Salix spp* contains many compounds that may act as root-promoting agents in cuttings along with fungicidal, insecticidal, and antibacterial properties [6]. Earlier, [7] found root-promoting substances in the extract of softwood cuttings of *S. alba*, and he confirmed that these substances synergistically interact with IAA to improve rooting in the mug bean cuttings. Other researchers concluded different results when applying water extract of *Salix spp* to the cuttings of different species. The best results were achieved in lavender (softwood) and chrysanthemum (semi-hardwood) cuttings at 1.06 μL.L$^{-1}$ willow bark extract [8]. Similarly, an extract of willow (*S. alba* L.) was successfully used as a natural rooting stimulator in mug bean cuttings [9]. However, 0.2% willow extract in gel form did not give a favorable rooting in stem cuttings of cannabis [10]. Besides, [11] observed that the extracts of *Salix babylonica* were not effective in increasing rooting in olive cuttings cv. 'Nabali'. Furthermore, *S. alba* is a deciduous tree, so phytochemicals and hormones in deciduous trees are variable throughout the year [12]. Tienaho *et al.* [13] reported that seasonal and environmental factors caused changes in phytochemical contents in the bark of *Salix spp*. Besides, [14] referred that during the vegetative season, precisely from March to July, secondary metabolites declined in the bark of willow clones.

As for propagation of thornless blackberries, they are readily propagated via clonal methods, digging of canes, crown division, dormant hardwood cuttings, softwood cuttings, leaf-bud cuttings, suckers, root cuttings, and layering [15]. Selection of one method relies on the number of new plants needed and the type of blackberry. In this regard, digging of canes, crown division, suckers, and layering do not allow the generation of new plants in large numbers for commercial production, because a high number of mother plants are required [16]. Additionally, root suckers and root cuttings are not recommended for the propagation of chimeral thornless cultivars due to losing thornlessness [17].

Cuttings propagation is widely used to propagate blackberries; however, the reports have revealed that the rooting of hardwood or floricane cuttings of blackberries are limited and variable. Nonetheless, cultivar, the time of taking cuttings, using auxins, and the cuttings atmosphere have decisive roles in the successful rooting of blackberry cuttings [18, 19]. In light of the above-mentioned reasons, in this study, willow shoots were collected at different times and extracted, and then applied to the cuttings of thornless blackberry in order to enhance rooting capacity.

**Materials and Methods**

The research was carried out in the College of Agricultural Engineering Sciences, University of Sulaimani, Kurdistan Region-Iraq to demonstrate the effect of willow shoot extract on the rooting of thornless blackberry cuttings as the shoots of willow were collected at different times (on the 15$^{th}$ of each month along the year).

**Collection of willow shoots and extractions**

The willow shoots (*Salix alba* L.) were collected from one- and two-year-old shoots on the 15th of each month in 2022, from a single tree. The shoots were cut into small pieces upon taking and dried in the shade, and then they were stored in plastic containers at room temperature until the time of extraction.

**Quantification of total phenols and flavonoids in willow shoots**

The dried willow shoots collected at different times were ground in a blender and used for analysis of total phenols and flavonoids according to [20].

**Colorimetric quantification of IAA in willow shoots**

Determination of endogenous IAA in willow shoots was conducted according to a slight modification in the methods described by [21, 22]. So, 1.5 g of dried and well-ground shoots of willow were mixed well with 10 mL of Salkowiski's reagent, shaken thoroughly, and left for 1 hr. After that, the mixture was filtered through a filter paper, and the absorbance was read at 530 nm. The concentration of IAA was found using the IAA standard curve and represented as $\mu g.mL^{-1}$.

**Determination of salicylic acid in willow shoots**

Salicylic acid was determined using a spectrophotometer as described by [23], with a slight modification. The samples were prepared by taking 0.05 g of dried and ground willow shoots, mixed with 1 mL ethanol at 20%, shaken for 20 minutes then centrifuged for 13 minutes at 13000 rpm. The supernatants were collected, and 100 µL was taken and mixed with 0.1% of $FeCl_3$ in 1% HCl up to 3 mL. After 30 minutes, the absorbance was read at 540 nm, and salicylic acid concentration was found using the salicylic acid standard curve as $\mu g.mL^{-1}$.

**Preparation of willow shoot extracts**

The extraction was initiated by grinding the dried willow shoots in a blender. The ground willow shoots which were taken from each collecting time (9 g) were extracted separately in (1 L) of 2% ethanol placed in a water bath at 35 °C for 3 hrs. and refrigerated for 24 hrs. After that, the extracts were filtered through filter papers, and the filtered extracts were applied to the cuttings.

**Taking and treatment of the cuttings**

The cuttings were taken on February 12, 2023 from the basal part of one-year-old shoots with 12 cm length (containing at least 3 buds) and 4-7.5 mm diameter. After transferring to the laboratory, they were soaked in 54% topsin fungicide at $1.5 \text{ mL.L}^{-1}$ for 30 min. The cuttings, after the surface drying, were soaked in the extracts of the willow shoots, 300 $mg.L^{-1}$ salicylic acid (SA) in 2% ethanol, 300 $mg.L^{-1}$ indole acetic acid (IAA) in 2% ethanol, and the control cuttings just in 2% ethanol for 1.5 hr. The treated cuttings were stuck in sand media prepared in black plastic pots as 7 cuttings per pot put inside a high tunnel. The experiment was laid out in randomized complete block design (RCBD) with three replications. The average minimum and maximum temperatures inside the high tunnel are shown in Figure (1).

**Statistical analysis**

The data were collected after 80 days (on May 3, 2023) from planting the cuttings. The measurements were rooting percentage, callusing percentage, death percentage, number of main roots, length of the longest root, shoot length, shoot diameter, leaf number, leaf area, chlorophylls *a* and *b* in the leaves, and survival percentage after transplanting. To calculate the survival percentage, the rooted cuttings were transplanted to a mixed media of fine sand + compost in polyethylene bags with a size of 8×20 cm, then they were placed in a lath-house for 45 days. The average minimum and maximum temperatures inside the lath-house was between (15.7- 33.4 ℃). Chlorophylls *a* and *b* were quantified according to [24]. The collected data were analyzed using XLSTAT software version 2019.2.2, one-way ANOVA-RCBD, and Duncan's multiple range tests (P ≤ 0.05) were used for comparison of the means.

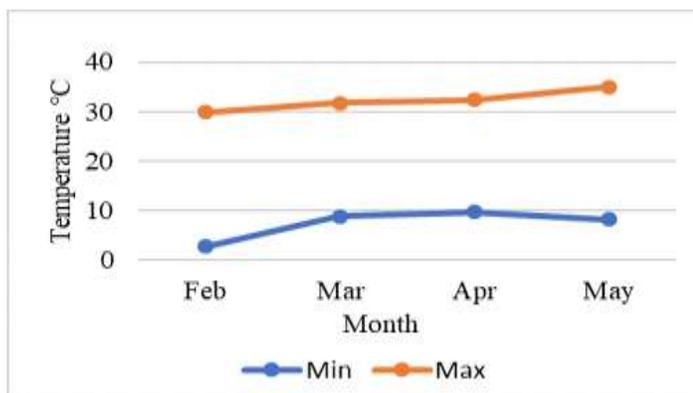

**Figure 1.** Average minimum and maximum temperatures inside the high tunnel, from February 12 till May 3, 2023.

**Results**

Analysis of the data showed that the extract of willow shoots collected on different dates gave different rooting%, callusing%, death%, and root length (Table 1), but collection dates of willow shoots showed no significant differences in rooting%, callusing%, death%, and root length related to control cuttings. The highest rooting percentage (66.67%) was detected in the cuttings treated with the extract of willow shoots collected on 15$^{th}$ of January followed by 15$^{th}$ of November (57.14%), and 15$^{th}$ of February (52.38%). In contrast, the lowest rooting percentage (33.33%) was observed in the cuttings treated with the extract of willow shoots collected on 15$^{th}$ of August and October accompanied by (38.09%) on 15$^{th}$ of April and June. Control cuttings in this study gave (42.85%) rooting. The callusing percentage was the highest (57.14%) in the treated cuttings with the extract of willow shoots collected on 15$^{th}$ of December, but it was the least (14.28%) in the treated cuttings with the extract of willow shoots collected on 15$^{th}$ of January. No dead cuttings were observed in the ones treated with the extract of willow shoots collected on 15$^{th}$ of December. Also, the willow shoot extraction taken from the collection date of 15$^{th}$ of November showed the lowest death (4.76%) of the thornless blackberry cuttings. While, the extract of the willow shoots collected on 15$^{th}$ of August resulted in the maximum number of dead cuttings (38.09%), followed by 15$^{th}$ of March (23.81%), and the treatment of 300 mg.L$^{-1}$ SA (23.80%). Additionally, the longest root (25.89 cm) was achieved from the cuttings supplied with the extract of willow shoots collected on 15$^{th}$ of October. At the same time, the extract of the willow shoots collected on 15$^{th}$ of September and February caused the shortest roots (13.46 and 13.79 cm, respectively).

The data shown in the same table confirm that soaking thornless blackberry cuttings in extracts of willow shoots collected on different dates made significant differences in comparison with control cuttings regarding root number and shoot length. Soaking the cuttings in 300 mg.L$^{-1}$ SA, the extract of willow shoots collected on 15$^{th}$ of December, and 300 mg.L$^{-1}$ IAA increased root number to the highest values (16.17, 14.8, and 14.67, respectively). Whereas, the extracts of willow shoots collected on 15$^{th}$ of March and July gave the minimum number of roots (4.33 and 4.67, respectively). Similarly, the collection of willow shoots on 15$^{th}$ of December and their extract led to the longest shoot (6.23 cm). Shoot length was the shortest (2.16, 2.52, and 2.59 cm) in the cuttings supplied with the extract of willow shoots collected on 15$^{th}$ of September and March, and in control cuttings, respectively.

Furthermore, shoot diameter was not significantly different in all treated and control cuttings in this study.

Table 1. Effect of 300 mg.L$^{-1}$ of salicylic acid (SA), 300 mg.L$^{-1}$ indole-3-acetic acid (IAA), and willow shoot extracts collected on different dates on rooting%, callusing%, death%, root number, root length, shoot length, and shoot diameter of thornless blackberry cuttings.

| Treatments | Rooting% | Callusing% | Death% | Root length (cm) | Root number | Shoot length (cm) | Shoot diameter (mm) |
|---|---|---|---|---|---|---|---|
| Control | 42.85 ab* | 42.85 ab | 14.28 ab | 17.55 ab | 6.93 cd | 2.59 cd | 1.98 a |
| SA 300 mg.L$^{-1}$ | 42.85 ab | 33.33 ab | 23.80 ab | 17.63 ab | 16.17 a | 4.5 a-d | 1.91 a |
| IAA 300 mg.L$^{-1}$ | 47.62 ab | 38.09 ab | 14.28 ab | 15.19 ab | 14.67 a | 2.61 cd | 1.85 a |
| 15$^{th}$ Jan. | 66.67 a | 14.28 b | 19.04 ab | 18.07 ab | 7.5 bcd | 3.43 bcd | 1.74 a |
| 15$^{th}$ Feb. | 52.38 ab | 38.09 ab | 9.52 ab | 13.79 b | 7.44 bcd | 3.92 a-d | 1.93 a |
| 15$^{th}$ Mar. | 42.85 ab | 33.33 ab | 23.81 ab | 15.11 ab | 4.33 d | 2.52 cd | 1.66 a |
| 15$^{th}$ Apr. | 38.09 b | 42.85 ab | 19.04 ab | 20.96 ab | 6.9 cd | 3.88 a-d | 1.7 a |
| 15$^{th}$ May | 47.61 ab | 42.85 ab | 9.52 ab | 15.31 ab | 8.53 bc | 5.65 ab | 1.87 a |
| 15$^{th}$ Jun. | 38.09 b | 42.85 ab | 19.05 ab | 17.03 ab | 7.62 bcd | 4.97 abc | 1.88 a |
| 15$^{th}$ Jul. | 47.61 ab | 33.3 ab | 18.95 ab | 15.34 ab | 4.67 d | 4.79 a-d | 1.77 a |
| 15$^{th}$ Aug. | 33.33 b | 28.57 ab | 38.09 a | 17.51 ab | 7.75 bcd | 3.72 a-d | 1.94 a |
| 15$^{th}$ Sep. | 47.62 ab | 42.85 ab | 9.52 ab | 13.46 b | 11.08 b | 2.16 d | 1.79 a |
| 15$^{th}$ Oct. | 33.33 b | 47.61 ab | 19.04 ab | 25.89 a | 6.37 cd | 5.9 ab | 1.95 a |
| 15$^{th}$ Nov. | 57.14 ab | 38.09 ab | 4.76 b | 17.17 ab | 9.87 bc | 4.35 a-d | 2.07 a |
| 15$^{th}$ Dec. | 42.85 ab | 57.14 a | 0 b | 23.78 ab | 14.8 a | 6.23 a | 2.02 a |

* Values in the same column taken the same letter were not significantly different subjected to Duncan's multiple-range test at ($p \leq 0.05$).

The evaluation of the variance demonstrated that the extracts of willow shoots which were collected on different dates had a significant role in improving leaf number, leaf area, and the ratio of chlorophyll *a* and *b* when they were compared with control cuttings (Table 2). Leaf number was plethora in the cuttings soaked in 300 mg.L$^{-1}$ SA, in the extract of willow shoots collected on 15$^{th}$ of January and December (8.08, 8.03, and 7.75, respectively). The cuttings soaked in the extract of willow shoots collected on 15$^{th}$ of September presented the least number of leaves (4.83). Further, willow extracts taken from the shoots collected on 15$^{th}$ of December and October exhibited the largest leaves (20.55 and 18.87 cm$^2$, respectively). Whereas, the smallest values of leaf area (10.07 and 10.76 cm) were recorded in the cuttings soaked in the extract of willow shoots collected on 15$^{th}$ of August and in control cuttings, respectively. Moreover, spectrophotometric analysis of leaf chlorophylls revealed that chlorophyll *a* (4.73 µg.g$^{-1}$) was the highest in the cuttings soaked in the extract of willow shoots collected on 15$^{th}$ of February, but it was the lowest (1.78 µg.g$^{-1}$) in the cuttings soaked in the extract of willow shoots collected on 15$^{th}$ of October. Besides, Chlorophyll *b* reached the maximum (3.09 µg.g$^{-1}$) in the cuttings soaked in the extract of willow shoots collected on the 15$^{th}$ of September, while minimum chlorophyll *b* (1.35 µg.g$^{-1}$) was determined in the leaves of the cuttings soaked in the extract of willow shoots collected on 15$^{th}$ of April. The results of survival percentage after transplanting, as indicated in table (2), also explained that there were no significant differences between the control cuttings and the cuttings that soaked in the extract of willow shoots collected in different dates. Besides, soaking the cuttings in 300 mg.L$^{-1}$ SA was the best to achieve (100%) survival after transplanting. However, the survival percentage after transplanting was the worst in the cuttings soaked in the extract of willow shoots collected on the 15$^{th}$ of May, June, July, August, and September, and it was more pronounced on the 15$^{th}$ of July and September (50%).

Table 2. Effect of 300 mg.L$^{-1}$ salicylic acid (SA), 300 mg.L$^{-1}$ indole-3-acetic acid (IAA), and willow shoot extracts collected on dates on leaf number, leaf area, chlorophylls *a*, *b* and Survival% after transplanting of thornless blackberry cuttings.

| Treatments | Leaf number | Leaf area (cm$^2$) | Chlorophyll *a* (μg.g$^{-1}$) | Chlorophyll *b* (μg.g$^{-1}$) | Survival% after transplanting |
|---|---|---|---|---|---|
| Control | 5.83 def* | 10.76 cd | 2.86 f | 1.83 g | 83.33 ab |
| SA 300 mg.L$^{-1}$ | 8.08 a | 13.78 a-d | 2.22 hi | 1.59 j | 100.00 a |
| IAA 300 mg.L$^{-1}$ | 5.16 ef | 12.34 bcd | 1.95 j | 1.64 i | 72.22 abc |
| 15$^{th}$ Jan. | 8.03 a | 13.63 a-d | 2.58 g | 1.73 h | 78.33 abc |
| 15$^{th}$ Feb. | 6.94 a-d | 15.89 a-d | 4.73 a | 2.87 b | 80.55 abc |
| 15$^{th}$ Mar. | 6.47 b-e | 11.18 bcd | 3.32 d | 1.85 g | 80.55 abc |
| 15$^{th}$ Apr. | 6.83 a-d | 16.51 a-d | 2.11 i | 1.35 l | 70.00 abc |
| 15$^{th}$ May | 6.2 c-f | 17.92 a-d | 2.66 g | 1.51 k | 58.89 bc |
| 15$^{th}$ Jun. | 7.17 a-d | 18.23 abc | 4.47 b | 2.72 c | 66.67 bc |
| 15$^{th}$ Jul. | 6.4 b-e | 14.86 a-d | 3.4 d | 2.07 e | 50.00 c |
| 15$^{th}$ Aug. | 5.94 def | 10.07 d | 2.34 h | 1.67 h | 66.67 bc |
| 15$^{th}$ Sep. | 4.83 f | 11.23 bcd | 2.96 ef | 3.09 a | 50.00 c |
| 15$^{th}$ Oct. | 7.5 abc | 18.87 ab | 1.78 k | 1.5 k | 91.67 ab |
| 15$^{th}$ Nov. | 6.61 a-e | 14.52 a-d | 3.01 e | 2.02 f | 69.44 abc |
| 15$^{th}$ Dec. | 7.75 ab | 20.55 a | 3.86 c | 2.37 d | 83.33 ab |

* Values in the same column taken the same letter were not significantly different subjected to Duncan's multiple-range test at (p ≤ 0.05).

Figure (2) illustrates that the quantities of total phenols, total flavonoids, salicylic acid (SA), and indole3-acetic acid (IAA) were apparently variable in willow shoots depending on the date on which the shoots were collected. In this regard, the collection of willow shoots on 15$^{th}$ of March and April brought about the maximal total phenols (57 and 56.27 μg.mL$^{-1}$) and SA (465.23 and 492.61 μg.mL$^{-1}$), respectively (Figure 2, A and C). Contrarily, minimal phenols (51.4, 51.7, 51.91, and 51.94 μg.mL$^{-1}$) were detected in the willow shoots collected on 15$^{th}$ of January, November, December, and August, respectively. In addition, the willow shoots collected on 15$^{th}$ of April had the peak value of total flavonoids (44.96 μg.mL$^{-1}$). Albeit, the least flavonoids were found in the willow shoots collected on 15$^{th}$ of January (29.07 μg.mL$^{-1}$) and 15$^{th}$ of August (29.18 μg.mL$^{-1}$). While, the collection of the willow shoots indicated that SA ratios were soaring in the willow shoots collected on 15$^{th}$ of April (492.61 μg.mL$^{-1}$) and March (465.23 μg.mL$^{-1}$), but the collection dates; 15$^{th}$ of October (197.55 μg.mL$^{-1}$) and January (228.57 μg.mL$^{-1}$) caused the lowest values. Further, IAA started to increase in the collected willow shoots from 15$^{th}$ of January until 15$^{th}$ of March (Figure 2D), and on 15$^{th}$ of March it was the highest (365.17 μg.mL$^{-1}$). Oppositely, IAA launched to decline from 15$^{th}$ of April to 15$^{th}$ of September, and it reached the minimum value (85.03 μg.mL$^{-1}$) on 15$^{th}$ of September. Once again, a slight increase in IAA was observed in the willow shoots collected on 15$^{th}$ of November (224.4 µg.mL$^{-1}$).

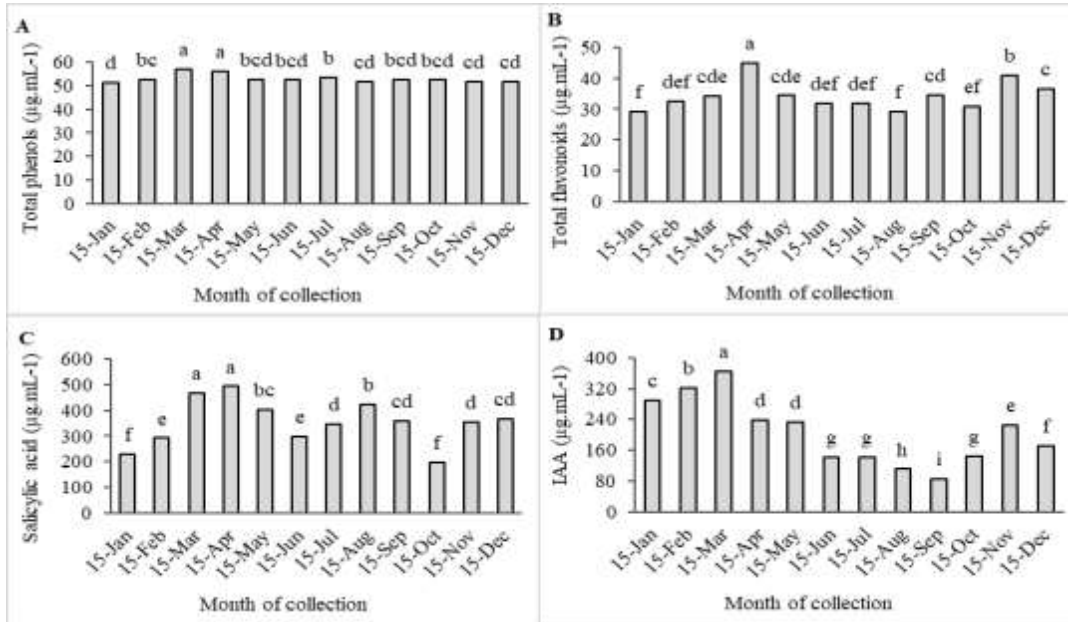

Figure 2. The concentration of total phenols (A), total flavonoids (B), salicylic acid (C), and IAA (D) in dried willow shoots collected on different dates. Columns with the same letter(s) show values which were not significantly different according to Duncan's multiple range test (P ≤ 0.05).

Pearson correlation test (P ≤ 0.05) of the means of the parameters is displayed in figure (3) showing that callusing percentage (Cal%) was negatively linked with death percentage (D%) (P=0.03), but it had a positive correlation with leaf area (LA) (P=0.03). Root length (RL) is positively associated with shoot length (SL), leaf number (LNo.), and leaf area (LA) (P=0.02, 0.03, and 0.03, respectively). Likewise, positive connections were found between SL and LNo. (P=0.02) as well as LA (P= 0.0001). Relationships between LNo. and LA (P=0.03), and LNo. and survival percentage after transplanting (SAT%) (P=0.01) were positive as well. In addition, chlorophyll *a* and chlorophyll *b* were positively related (P=0.0004)

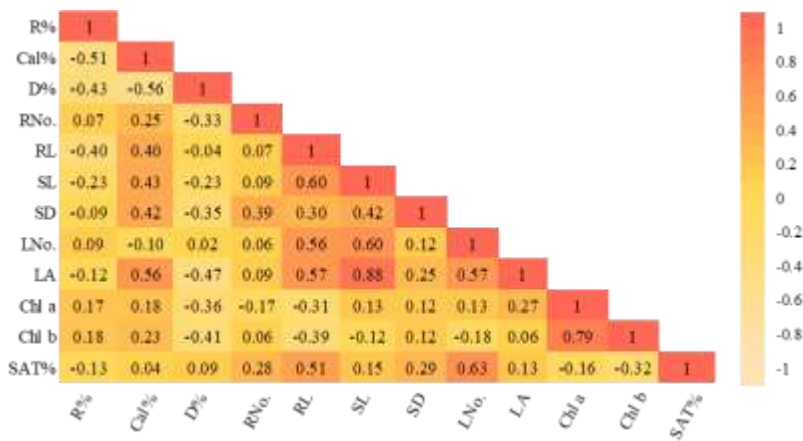

Figure 3. Interrelationships among studied parameters according to Pearson correlation test (P ≤ 0.05). Rooting percentage (R%), callusing percentage (Cal%), death percentage (D%), root number (RNo.), root length (RL), shoot length (SL), shoot diameter (SD), leaf number (LNo.), leaf area (LA), chlorophyll *a* (Chl a), chlorophyll *b* (Chl b), and survival percentage after transplanting (SAT%)

**Discussion**

The proportion of phytochemicals and hormones in willow shoots which were variable along the collection dates of the shoots might have a role in the regulation of rooting, callusing, and survival percentages of the thornless blackberry cuttings (Table 1 and Figure 2). In this context, the dates on which total phenols, total flavonoids, and/or salicylic acid (SA) were high in willow shoots reduced or kept rooting to a low extent, even if IAA was the highest. Also, low total phenols, total flavonoids, and/or SA in willow shoots did not significantly induce rooting in the cuttings of thornless blackberries if IAA was also low. On the contrary, low total phenols, total flavonoids, and/or SA and high IAA appeared to be favorable to improve rooting. So, the extract from the willow shoots collected on 15[th] of January displayed the highest rooted cuttings, at the meantime these willow shoots had the least total phenols, total flavonoids, and the second lowest SA, but they had the third highest IAA. In parallel, the lowest rooting percentage in the current study was found in the cuttings supplied with the extracts of the willow shoots collected on 15[th] of August and October. The willow shoots from the 15[th] of August collection contained low total phenols and total flavonoids, but the third highest SA, at the same time, IAA level existed in willow shoots was the second minimum. As for the willow shoots that were collected on 15[th] of October, they had the lowest SA and the third lowest total flavonoids, but IAA was also still low and total phenols were fairly high as well. However, IAA was the highest in the willow shoots collected on 15[th] of March, but their extract which did not improve rooting may be because of the highest total phenols and SA at the same time. There is a body of evidence that phenols may act as rooting promotors or inhibitors in cuttings [25, 26]. The type, concentration, and source of the phenols (endogenous or exogenous), and plant species have a decisive effect on the adventitious root formation (ARF) in the cuttings [27]. Some of the phenols inhibit IAA oxidation but others stimulate IAA oxidation [28]. IAA is outstanding for ARF in the cuttings. In the current study, total phenols were quantified in the willow shoots which may contain rooting inhibitor phenols in the collection dates that reduce rooting. On the other hand, different conclusions have been reached with the application of exogenous SA to the cuttings of different species. Therefore, at very low concentrations of 50 and 100 µM, SA helped ARF in cucumber hypocotyl cuttings by reducing the IAA conjugate levels [29]. Whereas, [30] found in olive cuttings that exogenous SA did not trigger any root formation, even though it was concluded that exogenous SA strongly halted rooting in comparison to IBA and NAA. The results of the death percentage in the present study verified that the cuttings soaked in the extract of the willow shoots collected on 15$^{th}$ of August presented the highest death percentage which included low total phenols, total flavonoids, and IAA but high SA. Death percentages were also high among the cuttings soaked in the willow shoots from the 15$^{th}$ of March collection, which contained high SA, and in the cuttings soaked in 300 mg.L$^{-1}$ SAIn spite of all of these, the phytochemicals and hormones in the willow shoots might enhance other characteristics of the thornless blackberry cuttings other than rooting, callusing, and death percentages. Thus, increasing root number due to 300 mg.L$^{-1}$ SA, 300 mg.L$^{-1}$ IAA, and the extract from willow shoots collected on 15$^{th}$ of December may belong to the fact that these treatments provided the cuttings with the necessities in a favorable concentration to rise root number in the rooted cuttings (Table 1). There are reports that SA and IAA influence some aspects of plant physiology processes in relation to elevating root numbers, such as cell division, antioxidant activities, and oxidative stress [31]. SA and IAA may be in an optimal balance with each other and other phytochemicals in the willow shoots collected on 15$^{th}$ of December for the best root number as well. High root lengths in the cuttings treated with the extract of the willow shoots collected on 15$^{th}$ of October, December, and August might be due to earlier rooting happened in these cuttings, therefore they had more time to elongate their root to the longest extent. Previous research indicated an association between earlier rooting and enhancing root length in the cuttings [32]. Besides, the phytochemicals and hormone profiles of these willow shoots were likely convenient for root elongation. In addition, the measurements revealed that the cuttings gave a better root system and concomitantly had a better shoot system (Tables 1 and 2). The thornless blackberry cuttings that possessed the best root number or root length simultaneously had the best shoot length, leaf number, and leaf area. Pearson correlation test demonstrated that a positive correlation was detected between root length and shoot length, and also between leaf number, and leaf area (Figure 3). Hence, the longest root might absorb higher water and nutrients needed for better shoot growth, in turn, the shoot system did more photosynthesis and sent more photosynthates to the root required for better growth. Shukla *et al.* [33] declared that for better root and shoot growth, a balance between root and shoot ratios is essential. Moreover, it is worth mentioning that the highest survival percentage after transplanting was achieved among the cuttings which endowed with the highest numbers of roots and leaves which were clear in the cuttings treated with 300 mg.L$^{-1}$ SA, and the second higher survival percentage was recorded among the cuttings that soaked in the willow extract form the shoots collected on 15$^{th}$ of October which had the best root length, shoot length, and leaf area. The best root and shoot characteristics are crucial for providing the cuttings with better resources required for surviving after transplanting. Survival percentage after transplanting is important because it determines the number of future new plants.

**Conclusion**

Based on the results obtained in this study, it can be deduced that willow shoot extracts made differences in the root number, shoot length, leaf number, leaf area, and chlorophyll *a* and *b* compared to control cuttings. The extracts of the willow shoots collected on different dates significantly affected the parameters of the thornless blackberry cuttings, some dates improved some characteristics of the cuttings and some dates decreased others, or were ineffective. Additionally, the phytochemical an d hormonal contents of the willow shoots were variable depending on the date of the collection. By the same token, it seemed that phytochemical and hormonal contents are related to the capacity of the willow shoot extracts to enhance the cutting characteristics. The best parameters measured when total phenols, total flavonoids, and/or SA were low and IAA was high, and this was prominent in the willow shoots collected on 15[th] of January, February, and November. However, further studies are needed to establish the outcomes of this study, particularly conducting specialized treatments to reduce the unfavorable consequences of the phenols, flavonoids, and/or SA in the willow shoots on the cuttings.

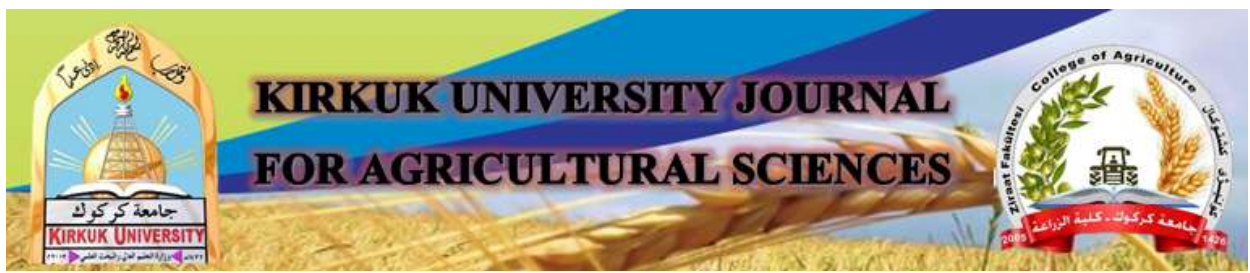

# تجذير اقلام توت العليق اللاشوكي المستحثة بمستخلص الصفصاف الأبيض (Salix alba L.) المجمعة في مواعد مختلفة

كوجر عمر صالح[1]    ئارام اكرم محمد[2]    ابراهيم معروف نوري[3]

kocher.salih@univsul.edu.iq    aram.hamarashed@univsul.edu.iq    ibrahim.nwri@univsul.edu.iq

[1], [2], [3] قسم البستنة، كلية علوم الهندسة الزراعية، جامعة السليمانية، السليمانية، إقليم كردستان، العراق.

- تاريخ استلام البحث 2024/3/5 وتاريخ قبوله 2024/3/24 .

## الخلاصة

يحتوي المستخلص المائي لنبات *Salix spp* على العديد من المركبات التي قد تعمل كعوامل تعزيز الجذور في العقل. *S. alba* هي شجرة متساقطة الأوراق التي تحتوي على العديد من المواد الكيميائية النباتية والهرمونات المتغايرة تختلف على مدار العام. لذلك في هذه الدراسة تم جمع الافروع *S. alba* بعمر سنة وسنتين في الخامس عشر من كل شهر في عام 2022، واستخلاصها في 2% إيثانول بمعدل 9 غ/لتر ووضعها في حمام مائي بدرجة حرارة 35 °C، ثم عوملت بها اقلام توت العليق اللاشوكي لمدة 1.5 ساعة. أوضحت النتائج أن أعلى نسبة تجذير (66.67%) تم الحصول عليها في العقل المغمورة بمستخلص افرع الصفصاف المجمعة بموعد 15 كانون الثاني التي لم تختلف معنوياً عن عقل المقارنة، لكنها كانت مختلفة عن العقل المغمورة في مستخلص افرع الصفصاف التي تم جمعها في 15 أب وتشرين الاول بنسبة (33.33%). كانت غالبية السمات الخضرية والجذرية الأخرى عالية في العقل المغمورة في مستخلص افرع الصفصاف التي تم جمعها في 15 ديسمبر. احتوت افرع الصفصاف التي تم جمعها في 15 كانون الثاني على أقل إجمالي للفينولات (51.4 ميكروغرام/مل) وإجمالي الفلافونويدات (29.07 ميكروغرام/مل). علاوة على ذلك، تم تسجيل أعلى إجمالي للفينولات (57 ميكروغرام/مل)، SA (492.61 ميكروغرام/مل)، وIAA (365.17 ميكروغرام/مل) في افرع الصفصاف التي تم جمعها في 15 ايار، وكان إجمالي مركبات الفلافونويد في أعلى مستواها (44.96 ميكروغرام/مل) في افرع الصفصاف التي تم جمعها في 15 نيسان. بشكل عام، بناءً على نسبة التجذير، يُنصح بجمع افرع الصفصاف في 15 كانون الثاني وشباط لاستخراجها وتطبيقها على اقلام توت العليق اللاشوكي.

**الكلمات المفتاحية**: الفلافونويدات، IAA، الفينولات، SA ، التجذير.